# Absorption spectrum of Gafchromic® EBT2 film with angular rotation

**Soah Park, Kwang-Ho Cheong, Taejin Hwang, Jai-Woong Yoon, Taejin Han,**

**Haeyoung Kim, Me-Yeon Lee, KyoungJu Kim, Hoonsik Bae, and Sei-Kwon Kang\***

*Department of Radiation Oncology, Hallym University College of Medicine, Seoul 150-950, Korea*

**Seung Jae Oh, Yong-Min Huh, and Jin-Suck Suh**

*Department of Radiology, College of Medicine, Yonsei University, Seoul 120-752, Korea*

It is important to study absorption spectrum in film dosimetry because the spectral absorbance of the film relates to the film's total absorption dose. We investigated the absorption spectra of Gafchromic EBT2 film with various rotational angles in a visible wavelength band. The film was irradiated with 6 MV photon beams and a total dose of 300 cGy. Absorption spectra were taken under different rotational angles after 24 h after irradiation and we fitted the spectra using Lorentzian functions. There were two dominant absorption peaks at approximately 586 nm (green) and 634 nm (red). The measured spectrum was decomposed 542 nm, 558 nm, 578 nm, 586 nm, 626 nm, 634 nm, and 641 nm. The maximum total area of the red band absorption spectrum was at 45° (225°) and the minimum at 90° (270°). As the angle of rotation changed, the intensity and integrated area of the blue and green peaks also changed with 180° period, with minima at 90° and 270°, and maxima at 0° and 180°, although the overall absorbance is very low. The spectral peak wavelengths remained constant within ±1.2 nm for all angles. There was no hysteresis of absorption spectrum of the film; spectra taken at 0° and 360° were substantially the same and showed similar behavior for all rotational angles. The change of absorbance with rotational angle of the film affected the dosimetric properties, resulting in rotational variations of film dosimetry in each red-green-blue channel.

PACS number: 87.53.Dq, 87.64.Ni, 87.53.Xd

Keywords: Gafchromic EBT2 film; Absorption spectrum; Rotational angle

Email: seikang@hallym.or.kr

Fax: +82-2-833-2839

# I. INTRODUCTION

Radiation treatment outcomes are related to the accuracy of the delivered dose to patients, so improvement in the precision of dosimeter and the accuracy of dose delivery are important to achieve successful radiation treatment. Moreover, dynamic dose delivery methods; such as Intensity-Modulated Radiation Therapy (IMRT), or Volumetric Modulated Arc Therapy (VMAT), need high accurate quality assurance in dose detection methods with high spatial resolution. Gafchromic film is a promising tool for precise and accurate two-dimensional dosimetry. It allows for clinically relevant doses to be accurately determined, provides excellent image resolution, low energy dependency [1, 2], near tissue equivalent properties [3], does not require chemical processing, and can be handled and prepared in room light and water. However, sources of uncertainty should be considered such as film rotation, edge effects, emulsion homogeneity, polarization, etc.

Absorption spectrum affects dosimetry characteristics because the delivered dose is related to the optical density. Hence, the optical properties of the film are important factors to understand film dosimetry in clinical radiotherapy applications. Mclaughlin et al. [4] reported that optical density (OD) measurements demonstrated that absorption peaks provide a sensitive form of dosimetry. Gafchromic EBT2 film includes a yellow dye in the radiation sensitivity layer and this produces a different physical construction of the substrate and overall thickness compared to previous versions of Gafchromic film. These may change light scattering effects in the film. Therefore, accurate knowledge of the film's optical properties is important for understanding the differences in dosimetric sensitivity of the film.

One of the important characteristics in Gafchromic film dosimetry is the spatial orientation of the film. Zeidan et al [5] found that the optical density varied by almost 50% by changing the orientation of the film. The active component in the EBT film is a needlelike polymer [6], and the polymers tend to align parallel to the coating direction of the film. The variation in optical density due to the alignment of polymers in the film produces a change in the spectral absorbance of the film. In this study, we investigate this change of absorption spectrum by rotating the orientation of Gafchromic EBT2 film.

## II. MATERIALS AND METHODS

The structure of Gafchromic EBT2 films (ISP Corporation, Wayne, NJ) [3] has an asymmetric layer configuration. The active layer contains a yellow marker dye that can serve as a reference level for film response and protect the active layer in the event of exposure to UV and visible light. The overall atomic composition of the EBT2 film is H (40.85%), C (42.43%), O (16.59%), N (0.01%), Li (0.1%), K (0.01%), Br (0.01%), and Cl (0.04%). We used a single film batch (lot A07160901) for this study. A film sheet was cut into 3 cm × 3 cm pieces. Absolute dose calibration of a linear accelerator (linac) (Varian 23iX LINAC, Varian Medical Systems, Palo Alto, CA) was carried out according to the IAEA-TRS398 protocol [7]. Film pieces were placed in a solid water phantom (RW3, PTW, Freiburg, Germany) with a thickness of at least 10 cm under the film to provide sufficient backscatter. The film pieces were positioned at the depth of maximum dose, with a source-to-surface distance (SSD) of 100 cm, and 10 cm × 10 cm beam field. Irradiation was performed with a 300 cGy dose using a 6 MV photon beam for absorption spectrum. The film's radiochromic analysis was delayed for at least 24 h to minimize the effects of post-irradiation coloration because post-exposure film darkening of 6%–9% within 24 h of irradiation has been reported in the EBT2 film [8].

We've obtained dose response curve to see film response to delivered dose of a 300 cGy. An Epson 1680 flatbed scanner (48-bit, Epson, Sydney, Australia) was used as the film scanning system [9]. The spatial resolution was 72 dots per inch (dpi). The films were scanned after several successive scans without film, because of the warming up effect of the scanner lamp [9]. The red channel was selected for dosimetry because the active layer maximum absorption peak is at 636 nm [10]. The scanned data are presented in terms of pixel values (PVs). We scanned un-irradiated films before irradiation in order to express in terms of net optical density (netOD) where the netOD is calculated as the difference in the optical densities of the exposed and the unexposed film [11].

The films were placed at the center of the scanner bed and scanned in the parallel direction. Image analysis was carried out using commercially available software, FilmQA (3cognition, Wayne, NJ). When a film is processed, FilmQA software applies a background correction for each film by using the scan of an unexposed film. We checked the repeatability of dose response of three films which were delivered a 300 cGy dose. Analysis was performed for the average regions of interest (ROI) consisting of $1 \times 1$ cm.

The absorption spectra were measured after 24 h of irradiation with a UV-1800 (SHIMADZU, Japan) UV/visible spectrophotometer. The spectrum was taken over the spectral range from 400 to 800 nm with spectral resolution of 1 nm, while the film orientation angle changed (0°, 45°, 90°, 135°, 180°, 225°, 270°, 315°, and 360°). 0 degree is parallel to the long dimension of the original $20.3 \times 25.4$ cm2 sheets (in Fig. 2). We obtained net absorption spectra [12] after subtracting background for all the angles respectively and fitted seven Lorentzian functions [13].

An absorption spectrum curve was fitted to the data using

$$y = y_0 + \sum_{i=1}^{7} \frac{2A_i}{\pi} \frac{w_i}{4(x - x_{ci})^2 + w_i^2}$$

, where $y_0$ represents baseline offset, $x_{ci}$ is center of given profile, $w_i$ is the full width at half maximum of the profile, and $A_i$ is total area under the curve from the baseline. The Levenberg-Marquardt (LM) algorithm, which is the most widely used algorithm in nonlinear least squares fitting, was used. We allowed fitting parameters vary. The lower boundary was $w_i > 0$ and we imposed no upper boundary.

### III. RESULTS AND DISCUSSION

Fig. 1 shows spectroscopic analysis for the net absorption spectrum of EBT2 film at a dose of 300 cGy when irradiated with 6 MV X-rays. The maximum dose variation between film and ionization chamber over three film measurements is 13 cGy, while the average dose deviation is 4.7 cGy. The measured spectrum was decomposed into Lorentzian functions and the combined fitting result is shown. Best fitting results were obtained when we used Lorentzian functions, with sharp peaks but long tails, such as are widely used in optical transitions between electron bands [13]. In our spectrum, the low wavelength peak around 430 nm had too low signal, so seven Lorentzian functions were used to fit the spectrum unlike Devic et al. [12], where 8 Lorentzian peaks were employed. $R^2$, representing the fit quality, was 0.9996. The reduced chi-square of the fit was $4.40 \times 10^{-5}$. Absorption spectra of EBT2 film [14] show a dominant peak in a red wavelength at a dose of 300 cGy. The decomposed peaks were 626 nm, 634 nm, and 641 nm in the red band. The peak at 634 nm showed the strongest absorption among these three Lorentzian peaks, as in Devic et al. [12]. The absorption maxima centered at 542 nm, 558 nm, 579 nm, and 634 nm. The peak in the green band was decomposed with 579 nm and 586 nm, using two Lorentzian functions because the shape of peak was asymmetric.

Fig. 2 shows absorption spectra with various orientation of the film. As the film was rotated, relative absorption peaks changed. The intensity of the peak in the red band increased toward 45° and decreased toward 135°. The peak in the green band decreases toward 90°, then increases toward 180°. The intensity of peak in the blue band is small (see Fig. 2), so it is difficult to compare the peaks in the region. However, the peak at 0° (black) and 180° (sky-blue) are both higher than that at 90° (green) and 270° (yellow).

The absorption spectra of the film at different rotational angles are compared in Fig. 3 ((a) 0°, (b) 45°, (c) 90°, and (d) 135°). The relative intensity of decomposed peaks changed under different rotational angles. For all angles, the average $R^2$ was 0.9996 and the standard deviation of $R^2$ was $8.95 \times 10^{-5}$. The average chi-square was $3.93 \times 10^{-5}$ and the standard deviation of chi-square was $1.42 \times 10^{-5}$. As the film rotated, the relative intensity of peaks changed. The average and standard deviation of the peak positions were 640 ± 0.5 nm, 633 ± 0.6 nm, 625 ± 1.2 nm, 586 ± 0.1 nm, 579 ± 0.2 nm, 558 ± 0.2 nm, and 542 ± 0.2 nm. The absorption peak positions were changed by rotating the film, but the effects were small. The change of absorbance of the film by rotating was large in the red and green bands, but small in the blue band. The relative absorption peak ratio also changed with rotation of the film.

Fig. 4 shows amplitudes of integral area of decomposed peaks of absorption spectra under different rotational angles. Peak 1 shows the highest variation in peak area, whereas peaks 2 and 3 are similar and somewhat reduced. The red peak, representing the sum of peaks 1–3 shows the same behavior as peak 1. The maximum peak absorbance is at 45° (225°) and minimum is at 135° (315°) for the red band. The variation of area of fitted peaks has 180° periodicity. The peak areas in the green and blue bands show different behavior on rotation. The maximum of absorbance for the green peak is at approximately 0° (180) and minimum at 90° (270°). Gafchromic EBT2 films also show a characteristic of birefringence [15]. Thus, its fast axis is at 45° (lowest refractive index) and its slow axis is at 135° (highest refractive index) [16]. The signal obtained from EBT2 film is strongly dependent on the red band wavelengths because the intensity of the absorption spectrum in the red wavelength is dominant. The largest absorbance is observed at 45° while the smallest one is at 135° for each 180° interval of film rotation. Devic et.al suggested that the blue part of the absorption spectrum would be beneficial to use when the film is irradiated to doses larger than 50 Gy.[17] Thus, the absorbance of EBT2 film can different with film orientation in high dose radiation, which can potentially enhance the absorption peak in the green or blue band preferentially.

## IV. CONCLUSION

The angular dependence of absorption spectra was observed by rotating the orientation of Gafchromic EBT2 film. The spectra were decomposed into seven Lorentzian functions, and the integral areas of the absorption peaks show periodicity of 180°. The maximum absorbance was at 45° and minimum at 135° in the red band, while the maximum was at 0° and minimum at 90° in the green or blue bands, each with periodicity of 180° rotation. It may be important to investigate absorbance further in the green or blue bands as well as the red band for higher radiation doses.

**ACKNOWLEDGEMENT**

This research was supported by Radiation Technology R&D Program through the National Research Foundation of Korea funded by the Ministry of Science, ICT & Future Planning (No.2013043498).

**REREFENCES**

[1] S. T. Chiu-Tsao, Y. Ho, R. Shankar, L. Wang, and L. B. Harrison, Med. Phys. **32**, 3350 (2005).

[2] M. J. Butson, P. K. N. Yu, T. Cheung, H. Alnawaf, Radiat. Meas. **45**, 836 (2010).

[3] International Specialty Products, http://www.filmqapro.com/Documents/GafChromic_EBT-2_20101007.pdf (2009)

[4] W. L. Mclaughlin, C. Yun-Dong, C. G. Soares, A. Miller, G. Van Dyk, and D. F. Lewis. Nucl Instrum. Methods. Phys. Res. A **302**, 165 (1991).

[5] O. A. Zeidan, S. A. Stephenson, S. L. Meeks, T. H. Wagner, T. R. Willoughby, P. A. Kupelian, and K. M. Langen, Med. Phys. **33**, 4064 (2006).

[6] A. Rink, I. A. Vitkin, and D. A. Jaffray, Med. Phys. **32**, 2510 (2005).

[7] IAEA Technical Reports Series No. 398. Vienna: International Atomic Energy Agency (2004).

[8] C. Andres, A. del Castillo, R. Tortosa, D. Alonso, and R. Barquero, Med. Phys. **37**, 6271 (2010).

[9] B. C. Ferreira, M. C. Lopes, and M. Capela, Phys. Med. Biol. **54**, 1073 (2009).

[10] S. Devic, N. Tomic, Z. Pang, J. Seuntjens, E. B. Podgorsak, and C. G. Soares, Med. Phys. **34**, 112 (2007).

[11] S. Devic, J. Seuntjens, E. Sham, E. Podgorsak, C. R. Schmidtlein, A. S.Kilrov, and C. G. Soares, Med. Phys. **32**, 2245(2005).

[12] S. Devic, S. Aldelaijan, H. Mohammed, N. Tomic, L. H. Liang, F. DeBlois, and J. Seuntjens, Med. Phys. **37**, 2207 (2010).

[13] A. P. Thorne, *Spectrophysics* (Champman and Hall, London, 1988).

[14] M. J. Butson, T. Cheung, and P. K. Yu, Australas. Phys. Eng. Sci. Med. **32**, 21 (2009).

[15] R. D. Guenther, *Modern Optics* (Wiley, NewYork, 1990).

[16] S. Park, S. K. Kang, K. H. Cheong, T. Hwang, H. Kim, T. Han, M. Y. Lee, K. Kim, H. Bae, H. Su Kim, et al, Med. Phys. **39**, 2524 (2012).

[17] S. Devic, N. Tomic, C. G. Soares, and E. B. Podgorsak, Med. Phys. **36**, 429 (2009).

Figure Captions.

Fig. 1. Fitting curves of the net absorbance for Gafchromic EBT2 film, which was irradiated with a dose of 300 cGy. The measured spectrum is shown in a solid black line, decomposed curves are shown in colored lines, and the combined fitted curve is presented by open circles.

Fig. 2. The optical absorption spectra for the EBT2 films with a dose of 300 cGy at different rotation angles.

Fig. 3. Fitting curves of absorption spectrum with rotational angles (a) 0°, (b) 45°, (c) 90°, and (d) 135°.

Fig. 4. The integral intensity of peak areas in red, green, and blue bands as a function of rotational angle. (a) Red, (b) Green, and (c) Blue represent the sum of the peaks in each wavelength band.

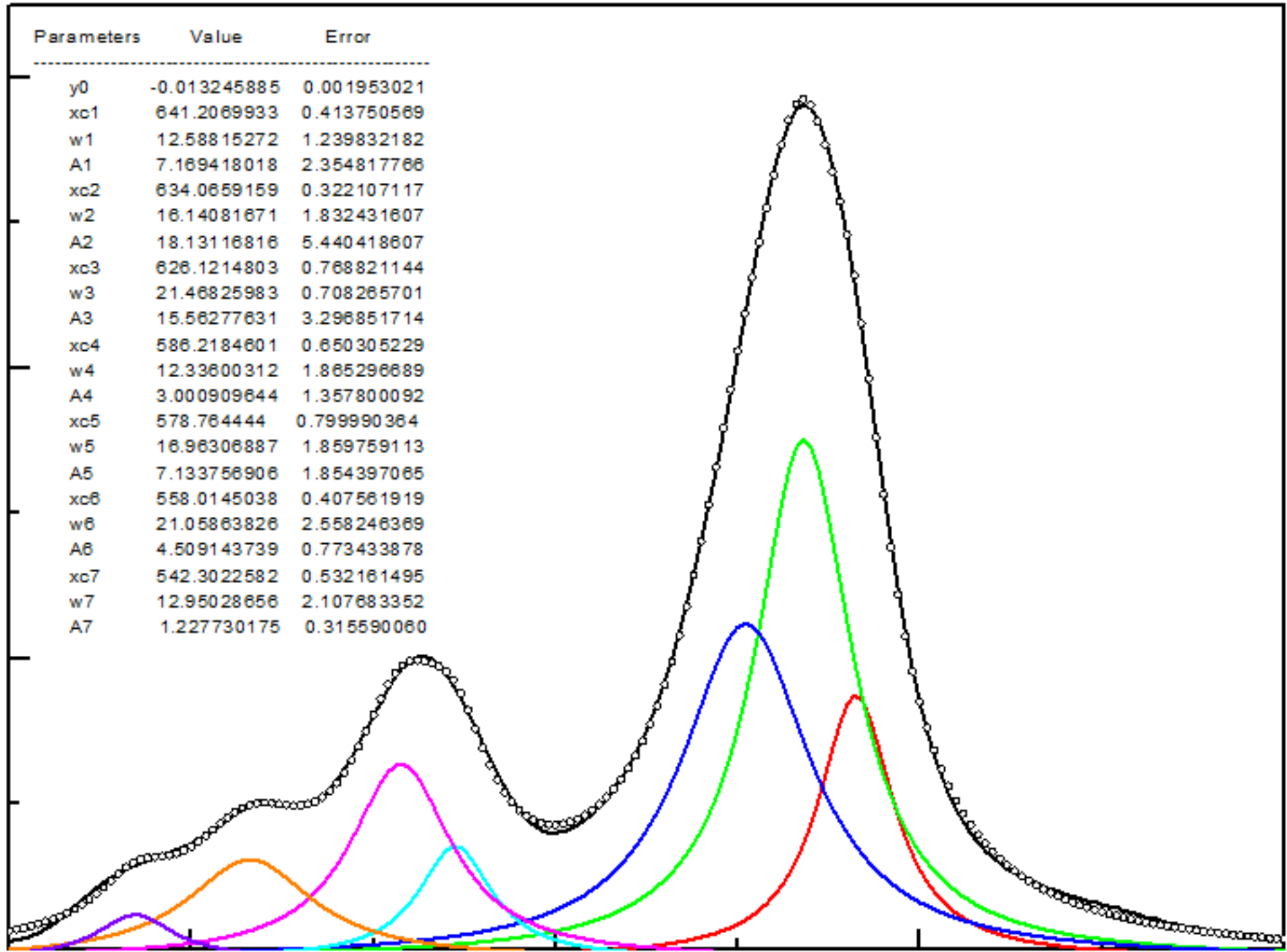

(a)

(b)

Measured spectrum
Peak1
Peak2
Peak3
Peak4
Peak5
Peak6
Peak7
Fitted Curve

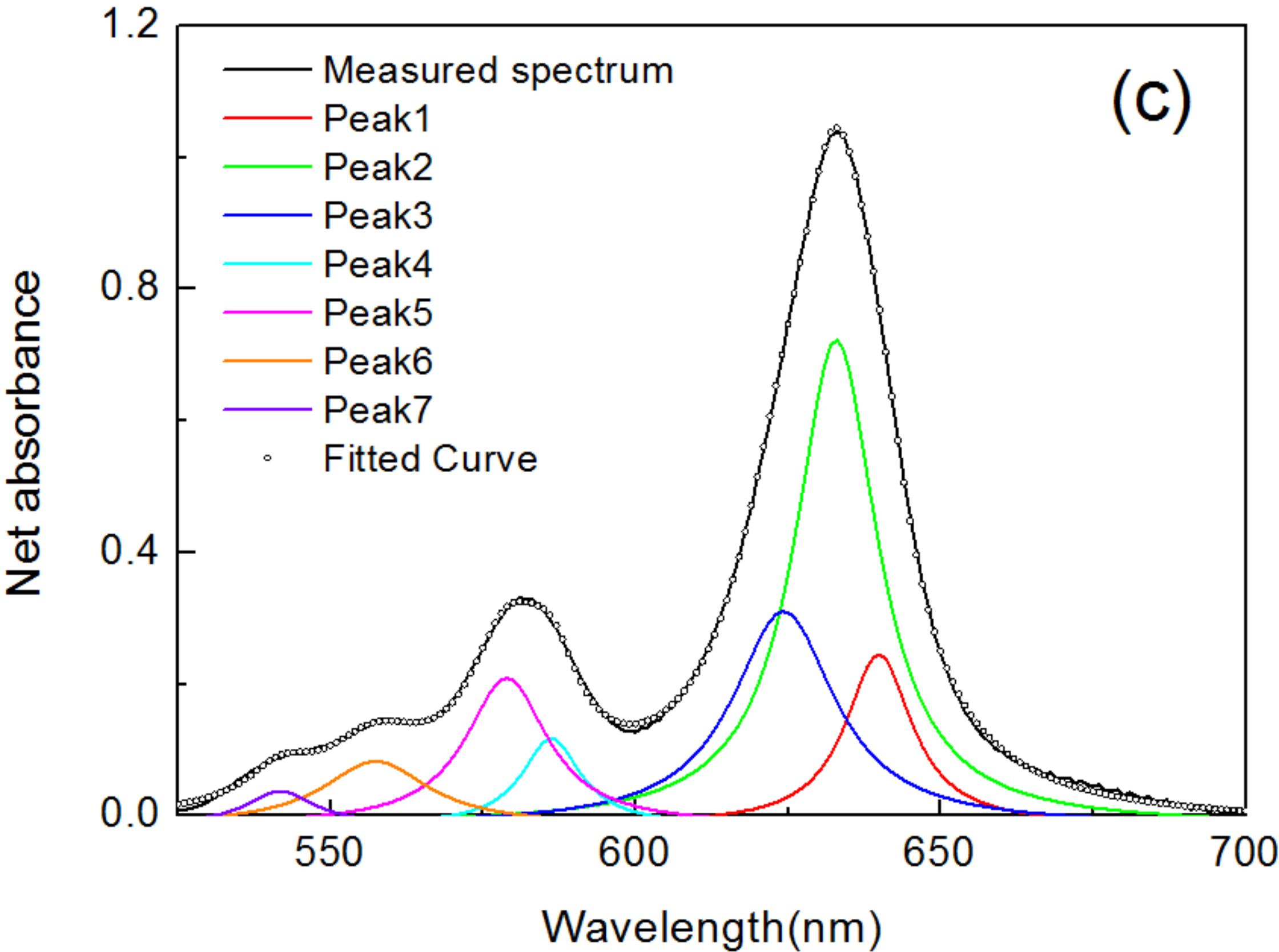

(d)

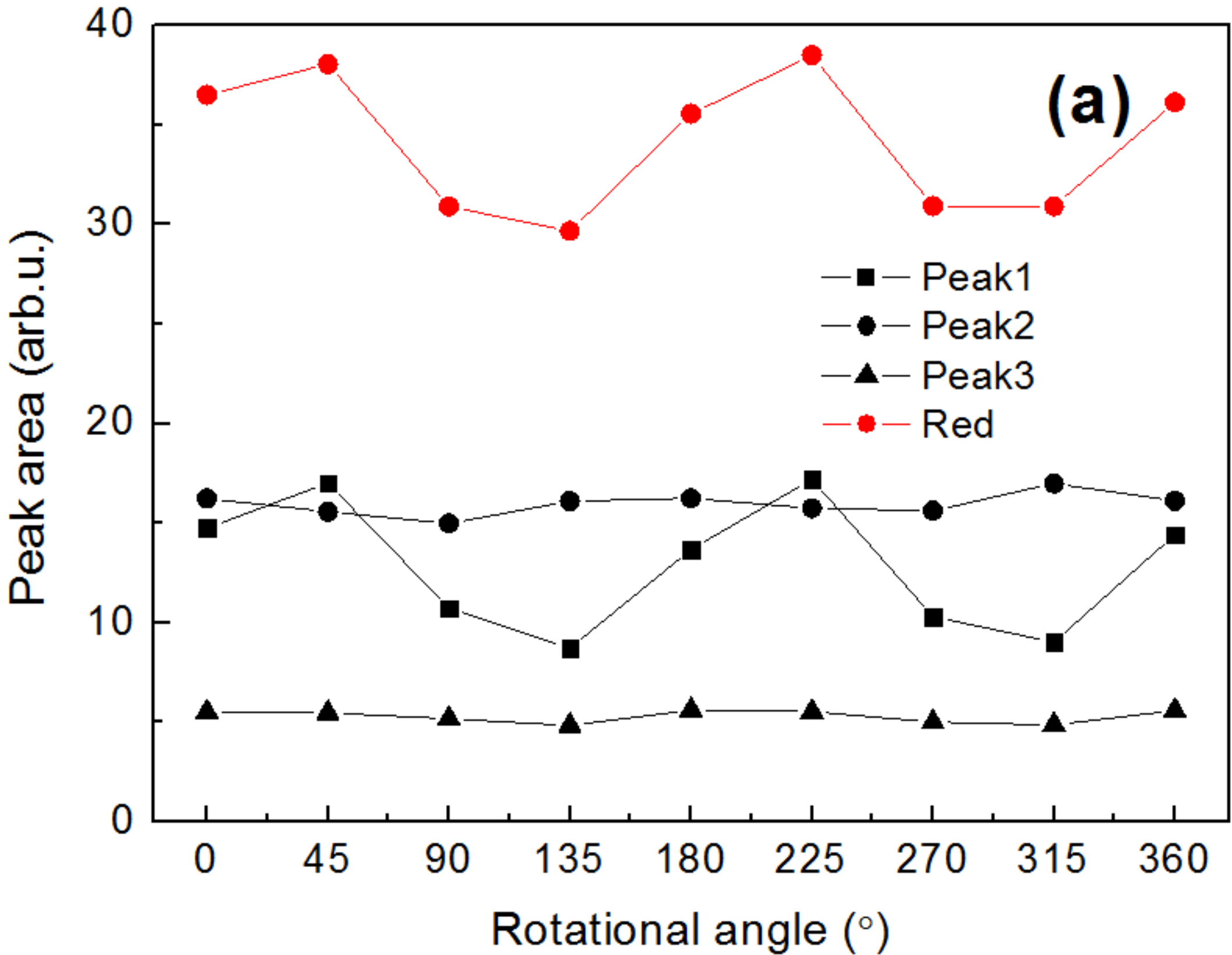

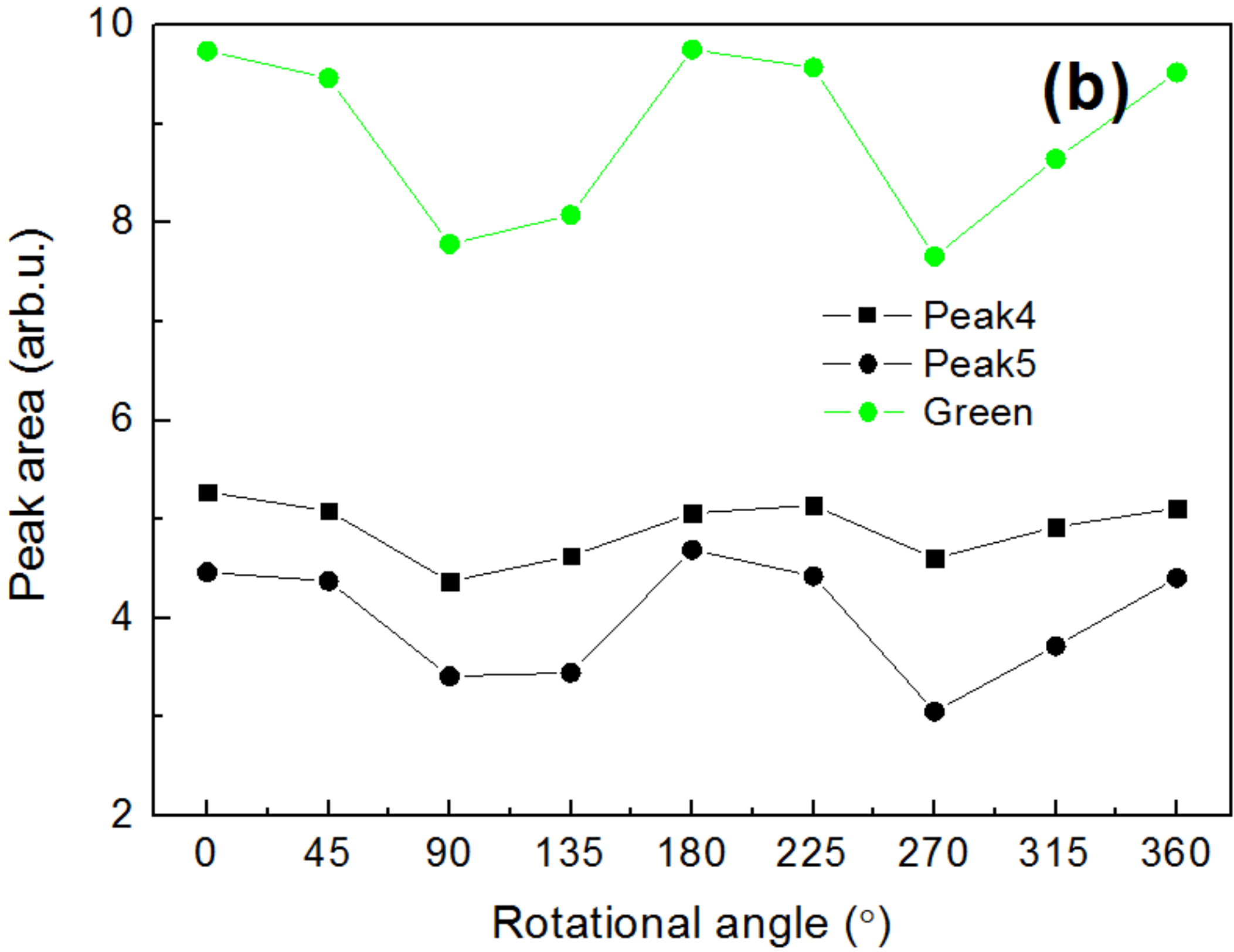

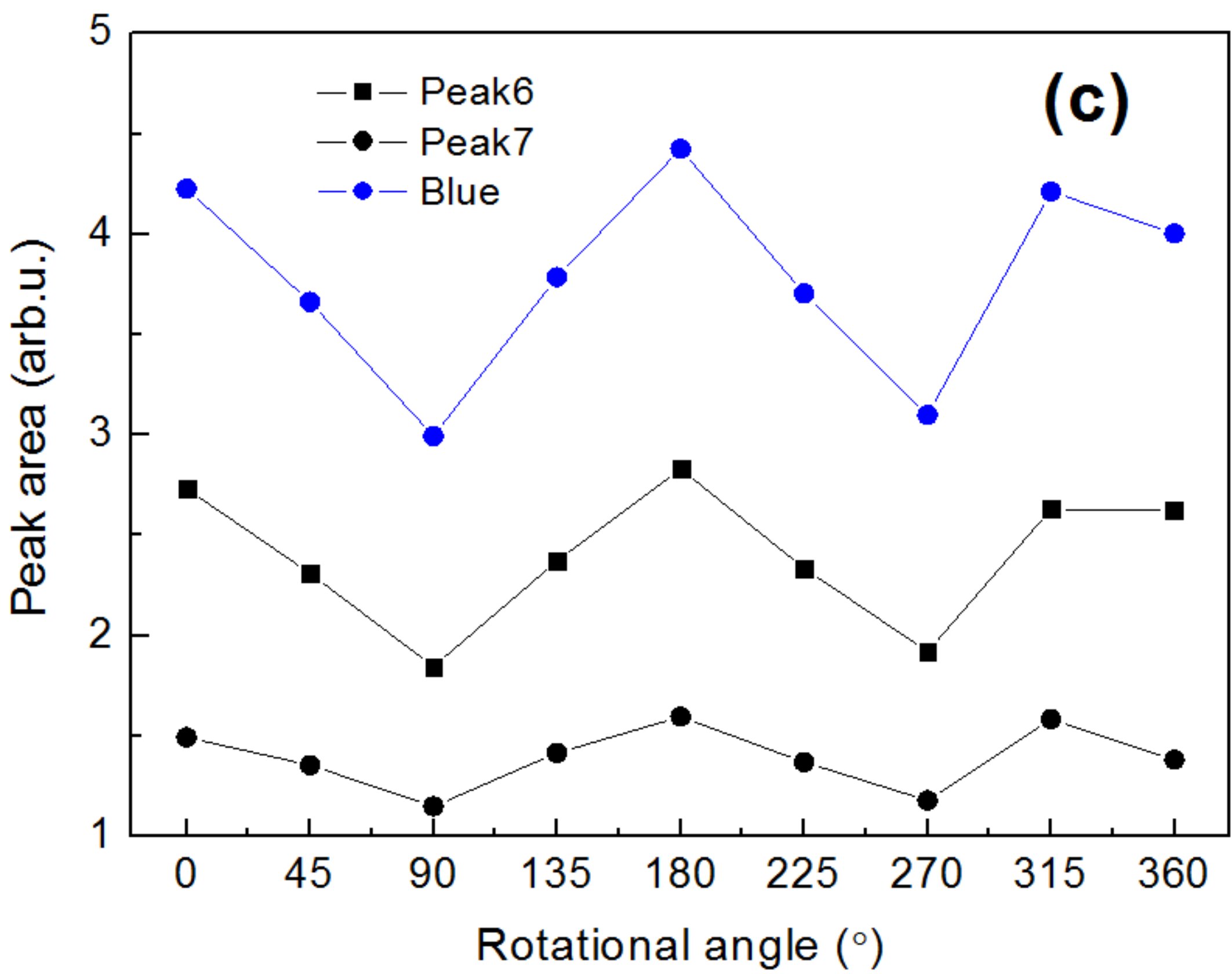